\documentclass[aps,prl,twocolumn,showpacs,groupedaddress]{revtex4}
\usepackage{graphicx}

\begin{document}


\title{Nucleation of Superconductivity in Thin Type-I Foils }



\author{P. Valko}
\altaffiliation{present address: Department of Physics, Slovak
Technical University, Ilkovicova 3, 81219 Bratislava, Slovakia}
\author{M.R. Gomes}
\altaffiliation{present address: University and INFN of Genoa, Via
Dodecaneso 33, 16146 Genoa, Italy}
\author{TA Girard}
\address{Centro de F\'isica Nuclear, Universidade de Lisboa, 1649-003 Lisbon, Portugal}


\date{\today}

\begin{abstract}
The expulsion of flux in five type-I materials in a slow,
continuously decreasing perpendicular magnetic field provides
evidence for the possible existence of a barrier in the
superconductive transition. The variation of the observed critical
fields with temperature yields Ginzburg-Landau parameter
determinations for the materials which suggests the behavior of the
study materials to be more strongly type-I than generally
considered.
\end{abstract}


\maketitle



The magnetic cycling of a type-I superconductor is fundamentally
hysteretic \cite{tinkh,degen}: the first order transition permits
superheating and supercooling states. For thin flat samples in a
perpendicular field, the hysteresis is even more pronounced because
of a demagnetization-generated, geometrical edge barrier
\cite{zeldov} which inhibits the penetration of flux in increasing
field. A topological hysteresis in the intermediate state flux
structures is also observed between crossing the phase line in
increasing or decreasing field \cite{huebener}.

It is commonly assumed that no similar barrier exists in decreasing
field \cite{brandt1,brandt2}, and that the expulsion of flux is
governed by the basic tenets of phase transitions. In the nucleation
regime ($H_{a} > H_{c2}$), only seeds of the superconductive phase
with size larger than a critical radius evolve; smaller seeds
collapse \cite{fud,liu}. In the spinodal regime ($H_{a} \leq
H_{c2}$), there is no free energy barrier to nucleation of the
superconducting phase and arbitrarily small seeds may evolve.  This
description however fails to treat the general nucleation of the
superconductive state during a continuous decrease of the applied
field. Neither does it include the effects of short- or long-range
interactions, nor effects associated with demagnetization or surface
nucleation.

Recent experiments on a tin foil in a continuously decreasing
applied field using a fast-pulse induction technique observed the
first expulsion of magnetic flux to occur at $H_{c3}$ \cite{jung},
which the authors then discounted as coincidental. We here report an
examination of the superconductive transition of several type-I
materials, listed in Table 1, at several temperatures in a gradually
decreasing magnetic field using fast-pulse techniques. The results
generally confirm $H_{c3}$ as the first flux expulsion field, and
suggest the existence of a barrier to the expulsion of flux. The
measured critical fields themselves moreover yield determinations of
the Ginzburg-Landau parameter $\kappa(T_{c})$ for the materials in
agreement with those obtained from measurements on superconducting
spheres, and a factor $\sim$ 2 below those derived from the more
accepted magnetization measurements on thin films/foils (which agree
with BCS estimates).

\begin{table}
  \caption{superconductive parameters of the study materials, from
  Ref. \cite{lide}, except for rhenium \cite{tulina}.}\label{Table 1}
  \begin{tabular}{|c|c|c|c|c|c|}
\hline  & $lead$ & $tantalum$ &  $rhenium$  &  $tin$ & $indium$ \\
\hline $\xi _{0} (\mu m)$ & $0.10$ & $0.097$ & $0.15$ & $0.23$ & $0.38$ \\
\hline $\lambda _{L} (\mu m)$ & $0.035$ & $0.032$ & $0.068$ & $0.034$ & $0.025$ \\
\hline $T_{c}$ (K) & $7.2$ & $4.5$ & $1.7$ & $3.7$ & $3.4$ \\
\hline $H_{c}$ (G) & $803$ & $829$ & $205$ & $305$ & $282$ \\
\hline
\end{tabular}
\end{table}

The fast pulse measurement technique has been described in detail
elsewhere \cite{jeudy1,jeudy3,jung}. The samples were cut from
$98.8-99.999\%$ pure, annealed, pinhole-free metallic foils of
10-125 $\mu$m thicknesses (d). Each foil was placed within a
rectangular copper pickup loop of 800 $\mu$m width, in contrast to
Ref. \cite{jung} where the tin strip was electroplated on only one
loop branch. The loop is transformer-bridged to a charge-sensitive
fast amplifier: a low frequency cutoff (10 kHz) on the bandwidth
prevents the recording of flux variations at the sweep rate of
$H_{a}$. Generally, only fast flux changes within the loop are
recorded: the nucleation of a flux bundle creates a discontinuity in
the flux intersecting the sense loop; the variation is a
$\delta$-function in time, and the response to the step variation is
obtained as long as the change is shorter than the nanosecond
risetime of the preamplifier. Imposition of a discriminator
threshold above the noise level defines the minimum recordable
amount of flux change, which we estimate at a few hundred
$\phi_{0}$. Extrapolation of the measurements to zero threshold
yields a noise-free determination of the characteristic transition
fields.

The measurements were performed in a single shot He3 refrigerator at
temperatures between 0.33 and 4.2 K, with an overall measurement
uncertainty of better than 0.5\%. The magnetic field was applied
perpendicularly to the sample by a coil external to the
refrigerator, with a homogeneity of 1\% over the sample area and
relative precision of better than $2 \times 10^{-4}$. The activation
of a gate is synchronized with the magnetic field step command so
that pulses originating on the pickup loop are recorded in the
appropriate field bin. Due to the large inductance of the magnet
coil, the signal is integrated in a linearly varying field; the
sweep rate was varied from a minimum of 0.5 Gauss/s to 250 Gauss/s,
although the results reported herein were systematically obtained
with a rate of $\leq$ 3 Gauss/s.

After zero-field cooling of the samples, measurements were performed
by recording all pulses above the discriminator threshold during
increase of $H_{a}$ at a constant rate from zero field to well above
the thermodynamic critical field $H_{c}(T)$ and subsequent return to
zero. Data were recorded separately for each direction of the field
sweep as a function of $H_{a}$. The data acquired during the field
increase were used to assess the foil quality and measurement
threshold level by monitoring the flux penetration profile, which in
the absence of noise yields a zero signal until the first
penetration field is reached.

A typical differential curve of the N $\rightarrow$ S transition
results is shown in Fig. 1. The abscissa is given in reduced applied
field $h_{a}(T) = H_{a}/H_{c}(T)$. The event count at each $h_{a}$
corresponds to a single field step decrease. The transition is
characterized by three regimes demarcated by two characteristic
fields. For the lowest thresholds, there is a characteristic first
expulsion field $H_{fe}$ indicated by a narrow signal, followed by
an absence of events for further decrease of $H_{a}$. This field
disappears with increasing threshold, suggesting it to consist of
small flux expulsions. There is in general no signal above $h_{fe}$,
except in cases where a direct correlation can be made with
perimeter metallurgical defects.

\begin{figure}[h]
\includegraphics[width=8 cm]{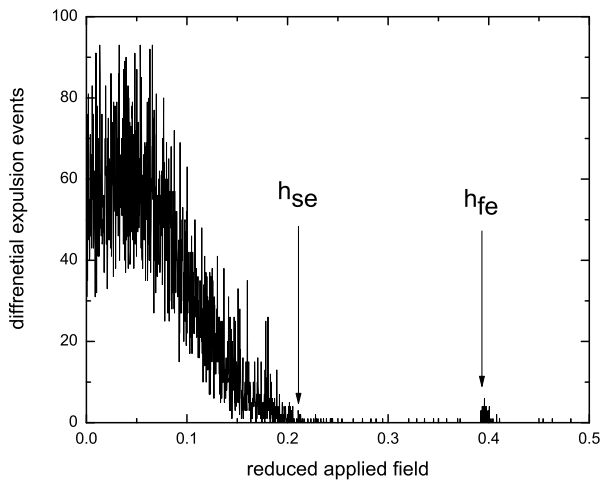}%
\caption{\label{}typical differential of the N$\rightarrow$S
transition signal obtained with a 50 $\mu m$ tin foil following zero
field cooling to 0.350 K and ramping of the applied magnetic field
to well above $H_{c}$.}
\end{figure}

The characteristic second expulsion field $H_{se}$ is indicated by a
rapid signal onset; this field persists with higher threshold
measurements, although the number of events below $h_{se}$ is
severely reduced. The largest amplitude pulses appear at the lowest
applied fields.

Similar transition curves were obtained with all materials studied,
for various aspect ratios and at different temperatures. Variation
of the strip positioning relative to the pickup loop, including
mounting the foil on a single branch of the loop, yielded no
qualitative differences at the level of experimental uncertainty.

Within uncertainties, there is typically no significant variation of
the characteristic fields with sample thickness, as shown in Fig. 2
for tin and rhenium. This identifies the two fields as intrinsic to
the materials.

\begin{figure}[h]
\includegraphics[width=7 cm]{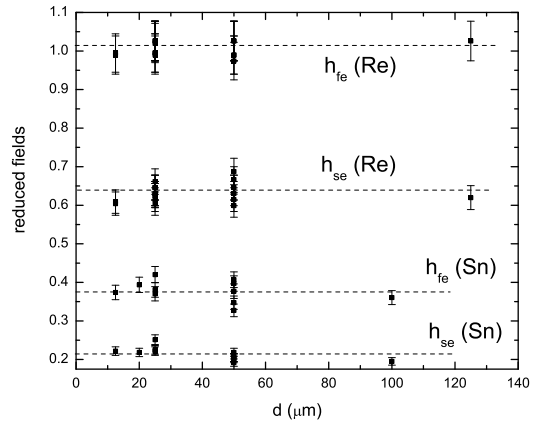}%
\caption{\label{}characteristic reduced N$\rightarrow$S transition
fields for different tin and rhenium sample thicknesses, at 350 mK.}
\end{figure}

For type-I materials, there are in general only two intrinsic fields
associated with the phase transition, $H_{c2}$ and $H_{c3}$. Near
$T_{c}$, $H_{c2}$ can be written as \cite{tinkh}

\begin{equation}\label{1}
H_{c2}(t) = \frac{\phi _{0}}{2\pi\xi ^{2}(t)},
\end{equation}

\noindent where $\phi _{0}$ is the flux quantum, $\xi (t) = \xi _{0}
(1-t)^{-1/2}$ and t=T/$T_{c}$ is the reduced temperature. In Fig. 3
we show the variation of $h_{se}$ with $1/(\xi ^{2}H_{c})$ for the
different materials and temperatures, assuming
$H_{c}(t)=H_{c}(0)[1-t^{2}]$. The lower line indicates the behavior
anticipated from Eq. (1) with $\xi _{0}$, $T_{c}$ taken from Table
I, and identifies $H_{se}$ with $H_{c2}$.

\begin{figure}[h]
\includegraphics[width=7 cm]{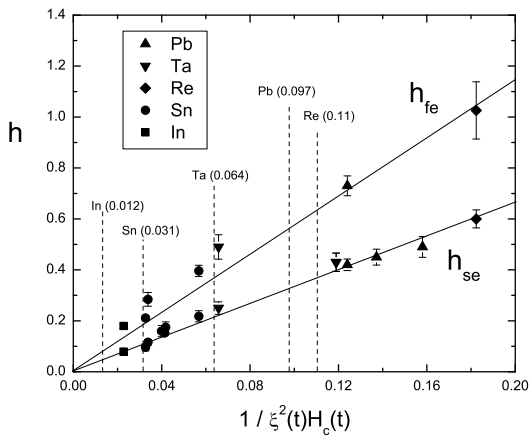}%
\caption{\label{} N$\rightarrow$S transition fields $h_{se}$ and
$h_{fe}$ for different materials and temperatures; the dotted lines
correspond to the abscissa at $T_{c}$, providing a lower limit to
each material's response.}
\end{figure}

If $H_{se}$ corresponds to $H_{c2}$, then Fig. 2 suggests that
$H_{fe} \sim 1.7H_{se}$ corresponds to $H_{c3}$. Fig. 3 also
displays $h_{fe}$ for the various materials, with the associated
line corresponding to a slope of 1.7$(\frac{\phi_{0}}{2\pi})$,
providing strong support for this identification.

First expulsion of flux at $H_{c3}$ without a complete collapse of
the normal state implies the spontaneous nucleation of
superconductivity in a surface sheath of width $\sim \xi(T)$ over at
least a part of the foil perimeter distant from the corners,
corresponding to the creation of a narrow flux-free band near the
foil edge. Although not associated with a barrier, such a band has
been observed in magneto-optic studies of Pb, Sn and In films
\cite{film,castro} in decreasing field. A similar band in increasing
field is commonly associated with the geometric barrier
\cite{zeldov,castro}, which separates the foil perimeter from the
intermediate state structure created by penetrated flux which is
driven to the minimum of the barrier potential near the foil center.
As indicated by Fig. 2, the band in decreasing field is not of
geometric origin.

With the existence of a perimeter band, further nucleation of
superconductive zones is technically impeded until the spinoidal
regime is reached and nucleation in the bulk of the foil becomes
feasible. The fact that signal below $H_{c2}$ is observed at all is
indication of a continuing barrier: flux expulsion occuring at the
rate of the field decrease alone is not observable with this
technique.

Additional indications of a barrier existence obtain from pauses
inserted at various $h_{a} < h_{c2}=$0.62 in the field ramp, shown
in Fig. 4; for $h_{a} > h_{c2}$, no signal is recorded. The vertical
lines are discontinuities between pause initiations and end, during
which signal was recorded in separate files (the event count of the
ramp resumption begins at the event number of the last pause event;
the small flat at the outset results from the response time of the
electronics); as seen in Fig. 4, this is exponentially-saturating in
time, and results from the decay of eddy currents in the magnet and
refrigerator.

A larger amplitude, identical response is observed (Fig. 4) with
pauses inserted in the penetration branch, which is generally
interpreted as the relaxation of the system to an equilibrium state
resulting from the lowering of the perimeter field by the
penetrating flux, re-raising the geometrical barrier
\cite{zeldov,castro}. Once equilibrium is established, no further
signal is recorded; further field increase is required to
re-initiate the penetration of flux. In the N $\rightarrow$ S
transition, the superconductive zones similarly continue to nucleate
following cessation of the downramp as a result of eddy currents,
with expulsion of the displaced flux, until an equilibrium is
established across the foil.

\begin{figure}[h]
\includegraphics[width=7 cm, height=2 cm]{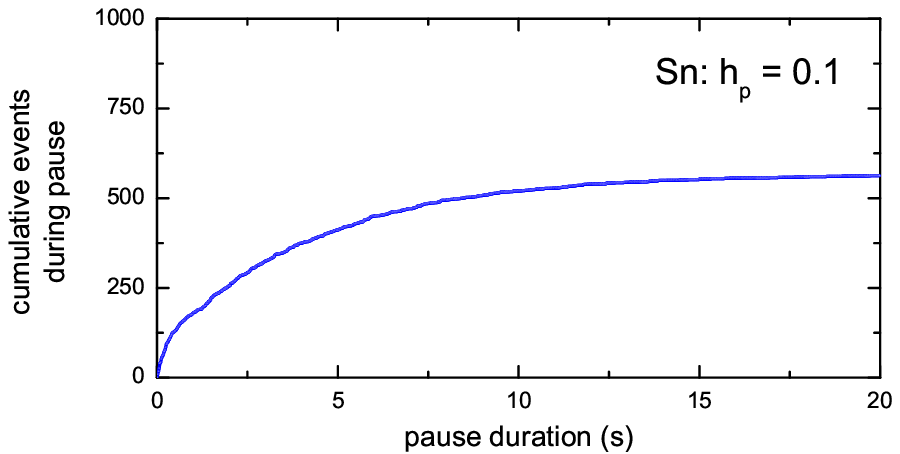}%
\end{figure}
\begin{figure}[h]
\includegraphics[width=7 cm]{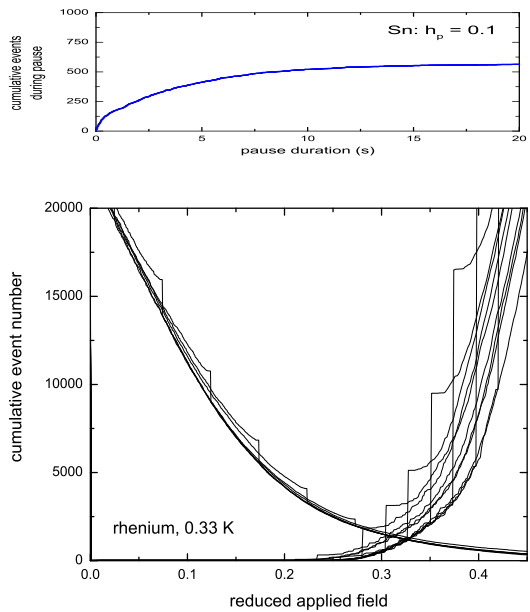}%
\caption{\label{}effect of a pause insertion at various fields in
field increases and decreases, for 25 $\mu$m rhenium at 330 mK. The
step results from flux changes occurring during the pause, which are
recorded separately as shown in the overfigure.}
\end{figure}


Observation of the critical fields permits an examination of the
Ginzburg-Landau classifications of the materials via $\kappa (t) =
(\surd 2)^{-1}h_{c2}(t)$ with $H_{c2} \rightarrow H_{se}$
\cite{degen}. The results of this analysis for all materials are
shown in Table II, in comparison with those previously extracted
(where available) from previous measurements on thin film/foil
\cite{chang,miller,auer}, microspheres \cite{feder,sbc}, and BCS
estimates based on Table I parameters. In those cases with
insufficient temperature measurements, the $\kappa$ has been
estimated from Fig. 3 based on the overall agreement of the results
with Eq. (1). Since $\xi ^{2}H_{c}$ = $\xi_{0} ^{2}H_{c}(0)[1+t]$,
there exists a lowest abscissa for each material (shown as dotted
vertical lines in Fig. 3) corresponding to $T_{c}$ which constitutes
a lower limit on $\kappa$. In all cases, the derived $\kappa$ are
consistent with those from the microsphere measurements
\cite{feder,sbc}, but significantly below the tabulations and thin
film/foil magnetization measurements.


The discrepancy is not related to field calibrations, as verified
using a triaxial Hall magnetometer; moreover, the first penetration
fields measured during field increase are in good agreement with
geometric barrier predictions \cite{benk}. The measured residual
resistivity ratios of the samples varied between 60-450, consistent
with impurities and lattice imperfections not playing a dominating
role in the results. These anyway would tend to decrease the
electron mean free path ($\langle l\rangle$), \textit{increasing}
all $\kappa$ by an additional $\kappa_{+} \sim
\frac{\lambda_{L}(0)}{\langle l\rangle}$. The results might also be
explained by an insufficient experimental sensitivity to small
amplitude pulses associated with smaller flux jumps at or below
noise level, except for the fact that the fields represent
zero-noise extrapolations.

The discrepancy in $\kappa$ between spheres and thin film/foil
determinations has been known for some decades, but to the best of
our knowledge remains unexplained. Curiously, the thin film/foil
results are in fact in better agreement with $\kappa$ derived from
$h_{c3}$ via $\kappa (t) = (1.695\surd 2)^{-1}h_{c3}(t)$, and also
agree in general with the lower temperature results of both the
spheres and this report. Curiously, the low temperature measurements
in rhenium indicate $H_{c2} < H_{c} < H_{c3}$, characteristic of
type 1$\frac{1}{2}$ materials ($\kappa \geq$ 0.42), despite the
determination of $\kappa_{Re}$ = 0.26$\pm$0.03 for which $H_{c2} <
H_{c3} < H_{c}$ \cite{tinkh,degen}. This suggests a variation of the
transition order with temperature, which has possibly important
ramifications since $\kappa$ is then less a fundamental property of
the superconductor than a simple ratio between the two
characteristic lengths in the description, both of which vary with
temperature and yield results consistent with the observed $\kappa$
determinations. Variation of the order of the transition with the
temperature is predicted in recent renormalization-based
reformulations of basic superconductive theory \cite{hove}, which
include fluctuations in the involved gauge and scalar fields, and
result in a dividing line between type-I and -II behavior at $\kappa
= 0.8/\sqrt{2}$ with a magnetic response which can be varied between
type-I and type-II simply by temperature change. This variation has
been seen in nitrogen-doped Ta ($\kappa$ = 0.665) \cite{auer}.

\begin{table}
  \caption{Survey of the Ginzburg-Landau parameters for the various
  materials. The tabulated $\kappa$ are obtained from the BCS $\kappa =
0.96\frac{\lambda_{L}(0)}{\xi_{0}}$ in the clean limit; the spheres,
from Ref. \cite{sbc}, the films/foils from Ref.
\cite{chang,miller}.}\label{Table II}
  \begin{tabular}{|c|c|c|c|c|c|}
\hline  $\kappa$ & $  lead  $ & $tantalum$ &  $rhenium$ &  $tin$ & $indium$ \\
\hline tabulated & $0.43$ & $0.32$ & $0.44$ & $0.16$ & $0.17$\\
\hline thin film/foil & $0.34$ & $0.36$ & $-$ & $0.15$ & $0.13$\\
\hline microspheres & $0.25(8)$ & $ - $ & $ - $ & $0.086(2)$ & $0.066(3)$\\
\hline this work & $0.23(3)$ & $0.15(2)$ & $0.26(3)$ & $0.09(1)$ & $0.036(4)$\\
\hline
\end{tabular}
\end{table}

The possible change in transition order with temperature in turn has
impact on current studies of quenched phase transitions in
superconductive systems as a means of obtaining information on the
formation of topological defects as seeds of large scale structure
in higher energy cosmological transitions \cite{zurek,kib2}. The
defect creation and distribution depends heavily on whether they
arise from gauge or scalar field fluctuations \cite{hindraj,kibraj},
and the order of the transition \cite{stephens,shap1,shap2}.

In summary, the nucleation of superconductivity in planar foils in
decreasing field, while characterized by the customary critical
fields of the phase transition, appears accompanied by a barrier to
the expulsion of magnetic flux. The observed critical field
variations with temperature further suggest the possible change in
the transition order with temperature. Given the implications of
these observations, further experiments to confirm or deny are
encouraged.

We thank G. Jung for informing us of the movie results of Ref.
\cite{castro}, and R. Prozorov and E. Brandt for useful discussions.
This work was supported by grants PRAXIS/10033/1998 and
POCTI/39067/2001 of the Foundation for Science \& Technology of
Portugal, co-financed by FEDER; it was also in part supported by
grant VEGA 1/2019/05 of the Slovak Republic. A part of this work was
accomplished within the context of the ESF program COSLAB.

\bibliography{conmatprl}

\end{document}